\documentclass[aps,prl,reprint,superscriptaddress]{revtex4-2} 
\usepackage[dvipsnames]{xcolor}
\usepackage[colorlinks=true,urlcolor=blue,citecolor=purple,linkcolor=red]{hyperref}
\usepackage{amssymb,amsmath,amsfonts}
\usepackage{epsfig}
\usepackage{graphicx}
\usepackage{epstopdf}
\usepackage{natbib}
\usepackage{braket}
\usepackage{dutchcal}
\usepackage{tikz,pgfplots}

\def\clock{{\count0=\time
		\divide\count0 60
		\ifnum\count0<10 0\fi\the\count0
		\multiply\count0 -60 \advance\count0 \time
		:\ifnum\count0<10 0\fi \the\count0
}}
\newcommand{\timestamp}{{\small\vbox{\hbox{\tt\jobname.tex}
			\hbox{\the\day/\the\month/\the\year, \clock}}}}

\hypersetup{%
	pdftitle   = {Strings near black holes are Carrollian},
	pdfkeywords = {Carrollian strings, black holes, non-Lorentzian geometry},
	pdfauthor  = {Arjun Bagchi, Aritra Banerjee, Emil Have, Jelle Hartong, Kedar Kolekar, Mangesh Mandlik},
}

\newcommand{\C}{\mathbb{C}}

\newcommand{\nn}{\nonumber}

\newcommand{\be}{\begin{eqnarray}}
	\newcommand{\ee}{\end{eqnarray}}
\newcommand{\beq}{\begin{eqnarray}}
	\newcommand{\eeq}{\end{eqnarray}}

\newcommand{\beqa}{\begin{eqnarray}}
	\newcommand{\eeqa}{\end{eqnarray}}
\newcommand{\D}{{\partial}}

\usepackage{mathrsfs}

\ifdefined\C
\renewcommand{\C}{\boldsymbol{C}}
\else
\newcommand{\C}{\boldsymbol{C}}
\fi

\definecolor{gris}{rgb}{0.5,0.5,0.5}
\definecolor{darkgreen}{rgb}{0.0,0.5,0.0}

\allowdisplaybreaks[0]
\usetikzlibrary{shapes.misc,shapes.symbols,snakes,decorations.text}
\tikzset{cross/.style={cross out, draw=black, thick, minimum size=2*(#1-\pgflinewidth), inner sep=0pt, outer sep=0pt},
	cross/.default={3pt}}
\pgfdeclarelayer{nodelayer}
\pgfdeclarelayer{edgelayer}
\pgfsetlayers{edgelayer,nodelayer,main}
\tikzstyle{ghost}=[fill=none, draw=none, shape=circle]
\tikzstyle{dot}=[fill=black, draw=black, shape=circle, scale=0.5]
\tikzstyle{grey line}=[-, draw={rgb,255: red,128; green,128; blue,128}]
\tikzstyle{blue line}=[-, draw=blue,thick]
\tikzstyle{red line}=[-, draw=red,thick]
\tikzstyle{blue line2}=[-, draw=blue]
\tikzstyle{dash blue}=[-, draw=blue, dashed]
\tikzstyle{dash red}=[-, draw=red, dashed, thick]
\tikzstyle{dash black}=[-, draw=black, dashed, thick]
\tikzstyle{dash grey}=[-, draw={rgb,255: red,128; green,128; blue,128}, dashed]
\tikzstyle{thick black line}=[-, thick]
\tikzstyle{thick black line dashed}=[-, thick,dashed]
\tikzstyle{blue fill}=[-, draw=none, fill={rgb,255: red,126; green,214; blue,255}]
\tikzstyle{black line}=[-]
\tikzstyle{thick red}=[-,draw=red,thick]
\tikzstyle{red fill}=[-, fill={rgb,255: red,255; green,162; blue,164}, draw={rgb,255: red,255; green,0; blue,4}]
\tikzstyle{purple fill}=[-, fill={rgb,255: red,128; green,0; blue,255}, draw={rgb,255: red,100; green,15; blue,128}]
\tikzstyle{arrow}=[draw=black, <-]
\tikzstyle{green arrow}=[draw=ForestGreen, ->]
\definecolor{colour1}{RGB}{252,57,0}
\definecolor{colour2}{RGB}{252,115,0}
\definecolor{colour3}{RGB}{252,173,0}
\definecolor{colour4}{RGB}{252,202,0}
\definecolor{colour5}{RGB}{252,255,130}
\definecolor{contournoin}{RGB}{255,255,0}
\definecolor{contournoout}{RGB}{228,0,5}

\begin{document}

	\title{Strings near black holes are Carrollian}

	\author{Arjun Bagchi}
	\email{abagchi@iitk.ac.in}
	\affiliation{Indian Institute of Technology Kanpur, Kanpur 208016, India}
	
	\author{Aritra Banerjee}
	\email{aritra.banerjee@pilani.bits-pilani.ac.in}
	\affiliation{Birla Institute of Technology and Science, Pilani Campus, Rajasthan 333031, India}

	\author{Jelle Hartong}
	\email{j.hartong@ed.ac.uk}
	\affiliation{School of Mathematics and Maxwell Institute for Mathematical Sciences,\\
		University of Edinburgh, Peter Guthrie Tait road, Edinburgh EH9 3FD, UK } 
	
	\author{Emil Have}
	\email{emil.have@nbi.ku.dk}
	\affiliation{Niels Bohr International Academy, Niels Bohr Institute, University of Copenhagen, Blegdamsvej 17, DK-2100 Copenhagen Ø, Denmark}
	
	\author{Kedar S.~Kolekar}
	\email{kedarsk@mail.tsinghua.edu.cn}
	\affiliation{Yau Mathematical Sciences Center, Tsinghua University, Beijing 100084, China}
	
	\author{Mangesh Mandlik}
	\email{mandlik@iitism.ac.in}
	\affiliation{Department of Physics, Indian Institute of Technology (Indian School of Mines) Dhanbad, Jharkhand 826004, India}

	\begin{abstract}
		We demonstrate that strings near the horizon of a Schwarzschild black hole, when viewed by a stationary observer at infinity, probe a string Carroll geometry, where the effective lightspeed is given by the distance from the horizon. We expand the Polyakov action in powers of this lightspeed to find a theory of Carrollian strings. We show that the string shrinks to a point to leading order near the horizon, which follows a null geodesic in a two-dimensional Rindler space. At the next-to-leading order the string oscillates in the embedding fields associated with the near-horizon two-sphere.
	\end{abstract}

	\maketitle
	\noindent \textbf{Introduction.} Classical general relativity comes together with the prediction of its own breakdown: spacetime singularities. The ultimate test of any theory of quantum gravity is the prediction of how such singularities are resolved. Since string theory is perhaps currently the best candidate of a theory of quantum gravity, questions of how strings perceive and resolve spacetime singularities are of fundamental importance. The cosmic censorship hypothesis \cite{Penrose:1969pc} says that singularities come hand-in-hand with horizons that shroud them. It is thus important to understand the behaviour of strings on black hole backgrounds.

	In this paper, we take important steps towards answering the question of how to formulate string theory near a black hole~\footnote{See, for example \cite{deVega:1987um,Lowe:1994ah,Bars:1994sv} for earlier attempts in this direction. This list is by no means exhaustive.}. For simplicity, we will focus our attention on the Schwarzschild black hole in four-dimensional (4d) asymptotically flat spacetimes. We will find that novel structures called {\em string Carroll geometries} play a starring role in our discussions. We expect that our construction naturally generalises to generic (non-extremal) black holes in arbitrary dimensions. More generally, Carrollian symmetries have been studied in the context of black holes in~\cite{Penna:2018gfx,Donnay:2019jiz, Hansen:2021fxi,perez:2021abf,Redondo-Yuste:2022czg} (see also~\cite{Ecker:2023uwm} for a construction of Carrollian black holes). 
	
	The near-horizon region of a non-extremal black hole, specifically the 4d Schwarzschild black hole, as viewed by a stationary observer at infinity, contains a 2d Rindler space. It has long been speculated that this 2d Rindler space plays an important role in the physics of the near-horizon region \footnote{See for example \cite{Townsend:1997ku} for a detailed introduction.}. The near-horizon geometry is the product of 2d Rindler space and a $2$-sphere, and in this region distinct points on the 2-sphere become causally disconnected, making it unclear how to proceed systematically. We show below that this near-horizon region is described by a string Carroll expansion of the Schwarzschild metric, borrowing nomenclature from its Galilean cousin, the string Newton--Cartan expansion~\cite{  Andringa:2012uz,Bergshoeff:2019pij}.

\begin{figure}[ht!]
		\centering
		\includegraphics[width=0.5\textwidth]{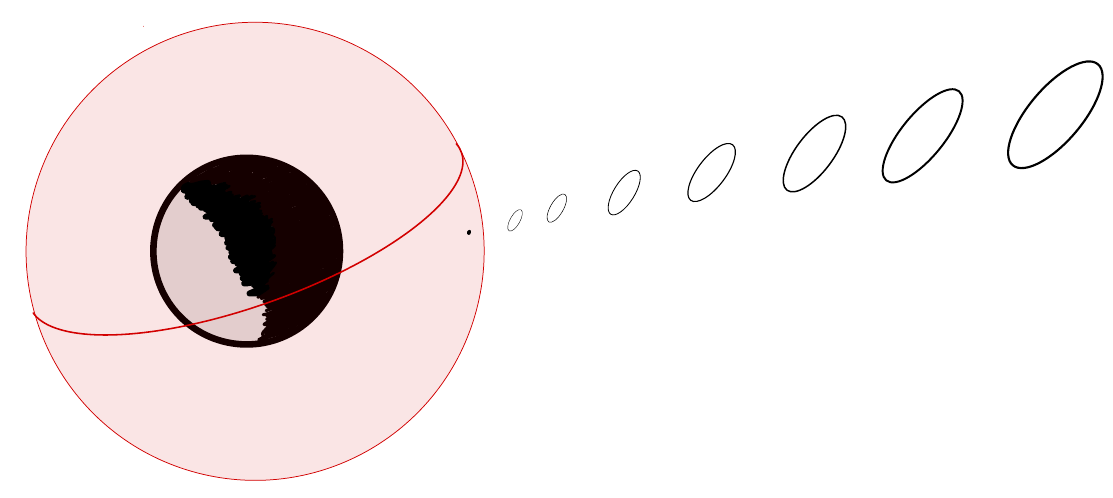}
		\begin{tikzpicture}[overlay]
			\begin{pgfonlayer}{nodelayer}
				\node [style=ghost] (0) at (2, 1) {{$\epsilon\rightarrow 0$}};
				\node [style=ghost] (2) at (-2.2, 0.26) {$r=r_h$};
				\draw [densely dotted,<-] (-0.2,1.3) to (4,1.3);
			\end{pgfonlayer}
		\end{tikzpicture}
		\caption{An asymptotic observer sees a closed string shrinking to a point as it approaches the near-horizon region, which looks like 2d Rindler space times a 2-sphere. In this regime, the string effectively probes a string Carroll geometry. The distance from the horizon $\epsilon$ provides an effective speed of light.}
		\label{fig:NH-Carroll-string}
	\end{figure}
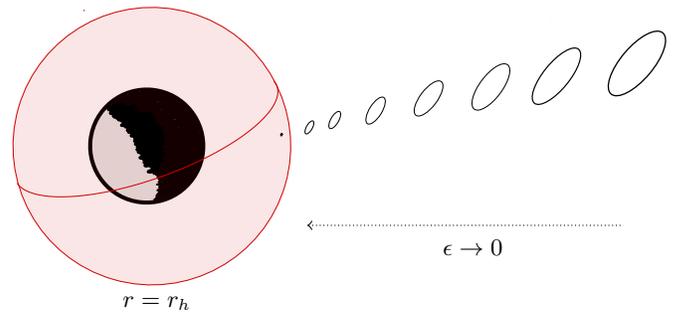

	In Schwarzschild coordinates the distance between two points on the sphere versus the distance between two points in Rindler space gets larger and larger as one approaches the horizon. Hence, strings probing the near-horizon region of a 4d Schwarzschild black hole are a priori confronted with very large spherical dimensions and a very small 2d Rindler space. So from this perspective, it seems that the string would principally see the sphere, rendering the 2d Rindler space unimportant. Through a careful analysis of the string Carroll expansion of the Polyakov action, we will show that the string ``freezes'' at leading order on the sphere and hence the 2d Rindler space does 
	become important, cf.~Figure~\ref{fig:NH-Carroll-string}. We will study solutions of the equations of motion of this string in the near-horizon region and show that the string behaves like a particle following a null geodesic in the 2d Rindler space while retaining subleading transverse wave-like properties on the sphere. 
	
	A comment here for the Carroll aficionado: since the velocity decouples from the momentum, Carrollian theories are best formulated in phase space. Starting from a relativistic phase space theory, there are two inequivalent Carrollian limits leading to what are generally called the electric and the magnetic theories. The limit of the Polyakov string that we consider here can be rephrased in terms of a phase space action corresponding to the magnetic theory. The phase space analysis, including details of generic electric and magnetic Carroll strings as well as further applications in the context of strings near black hole horizons, will be considered in upcoming work~\cite{UpcomingCarrollString}.  
	
	\medskip
	
	\noindent \textbf{Carrollian structures and black holes.}
	The Carroll algebra arises as the speed of light $c \to 0$ limit of the Poincare algebra \cite{NDS, LBLL} and has recently been observed in a wide array of physical systems ranging from holography of asymptotically flat spacetimes, where the conformal version of the Carroll algebra appears as the symmetries of the putative boundary dual theory \cite{Bagchi:2010zz, Barnich:2012aw, Bagchi:2012xr,Barnich:2012xq,Bagchi:2012cy,Bagchi:2014iea,Hartong:2015usd, Bagchi:2016bcd,Donnay:2022aba,Bagchi:2022emh,Figueroa-Ofarrill:2021sxz,Donnay:2022wvx, Bagchi:2023fbj, Saha:2023hsl, Bagchi:2023cen, Mason:2023mti}, to applications in cosmology \cite{deBoer:2021jej}, fluid models for the quark-gluon-plasma \cite{Bagchi:2023ysc, Bagchi:2023rwd} (more generally, see~\cite{Ciambelli:2018wre,Redondo-Yuste:2022czg, Ciambelli:2018xat, Campoleoni:2018ltl, Petkou:2022bmz, Freidel:2022bai, Freidel:2022vjq, deBoer:2023fnj,Armas:2023dcz} for Carrollian hydrodynamics) and condensed matter systems in the context of fractons \cite{Bidussi:2021nmp,Marsot:2022imf, Figueroa-OFarrill:2023vbj,Figueroa-OFarrill:2023qty} and flat bands \cite{Bagchi:2022eui}. In string theory, the very high energy limit where the string tension goes to zero is governed by conformal Carrollian symmetry arising on the string worldsheet replacing Virasoro symmetries \cite{Isberg:1993av, Bagchi:2013bga, Bagchi:2015nca}. Recently, Carrollian string theories on $p$-brane Carrollian geometries, which generalise the string Carroll geometry by having $(p+1)$ longitudinal directions, appeared in a duality web of decoupling limits of string theory in~\cite{Blair:2023noj,Gomis:2023eav}. 
	
	Moreover, Carrollian symmetry is known to arise on generic null hypersurfaces \cite{Hartong:2015xda}, and therefore black hole horizons in particular come equipped with a Carrollian structure~\cite{Donnay:2019jiz} (see also~\cite{Fontanella:2022gyt} for a discussion of non-Lorentzian characteristics of near-horizon geometries). The null Raychoudhuri and Damour equations of black holes can be reinterpreted as Carroll hydrodynamic equations: here the distance from the horizon acts as an effective speed of light. 
	
	Below we build on this notion and show that not only is the horizon Carrollian, the near-horizon region arranges itself as a generalisation of a Carrollian expansion, where we have two instead of one distinguished directions (spanning a Carrollian lightcone structure where every point on the near-horizon 2-sphere is causally disconnected), i.e., we are dealing with a string Carroll expansion of the near-horizon metric. 
	
	\medskip
	
	\noindent \textbf{Near-horizon region admits a string Carroll expansion.} In the non-relativistic regime stringy versions of Newton--Cartan geometry  \cite{Andringa:2012uz,Bergshoeff:2019pij} are well-known. For the Carroll case this is less so and we will next consider their properties. We will follow the approach to string Newton--Cartan geometry of~\cite{Hartong:2021ekg,Hartong:2022dsx} (as well as the $c=\infty$ and $c=0$ expansion approach of \cite{Hansen:2018ofj,Hansen:2020pqs,Hansen:2021fxi,deBoer:2023fnj}).  In order to construct a $D=(d+2)$-dimensional string Carroll geometry from a Lorentzian geometry of the same dimension we start by writing the metric and its inverse as 
	\begin{equation*}
		g_{\mu\nu} = c^2\eta_{AB}T_\mu^A T_\nu^A + \Pi^\perp_{\mu\nu}\,,~~g^{\mu\nu} = c^{-2} \eta^{AB} V^\mu_A V^\nu_B + \Pi^{\perp\mu\nu}\,,
	\end{equation*}
	where $c$ is the speed of light, and where $\mu,\nu=0,1,2,3$ are spacetime indices, while $A,B=0,1$ are longitudinal tangent space indices. The tensor $\Pi^\perp_{\mu\nu}$ has signature $(0,0,1,1)$. By construction, these objects satisfy the relations $V^\mu_A T_\mu^B = \delta^A_B$ and $T_\mu^A\Pi^{\perp\mu\nu} = V^\mu_A \Pi^\perp_{\mu\nu} =0$. The string Carroll expansion of these objects in powers of $c^2$ takes the form
	\begin{equation}
		T^A_\mu = \tau^A_\mu + \mathcal{O}(c^2)\,,~~\Pi^\perp_{\mu\nu} = h_{\mu\nu} + c^2\Phi_{\mu\nu} + \mathcal{O}(c^4)\,,
	\end{equation}
	with $V^\mu_A$ and $\Pi^{\perp\mu\nu}$ expanding in a similar way. Combining these expressions, the metric becomes
	\begin{equation}
		\label{eq:carroll-exp}
		g_{\mu\nu} = h_{\mu\nu} + c^2 \eta_{AB}\tau^A_\mu \tau^B_\nu + c^2 \Phi_{\mu\nu} + \mathcal{O}(c^4)\,.
	\end{equation}
	In a string Carroll geometry the tangent space has longitudinal directions corresponding to the vielbeine $\tau^A_\mu$ and transverse directions corresponding to the nonzero eigenvalues of $h_{\mu\nu}$ which has signature $(0,0,1,1)$. The Lorentz boost transformation that acts on $T^A_\mu$ and $\Pi^\perp_{\mu\nu}$ contracts to a string Carroll boost transformation acting on $\tau^A_\mu$ and $\Phi_{\mu\nu}$ while leaving $h_{\mu\nu}$ and $\eta_{AB}\tau^A_\mu \tau^B_\nu +  \Phi_{\mu\nu}$ inert. We refer to the upcoming paper~\cite{UpcomingCarrollString} for more details.
	
	Consider now a $D=4$-dimensional Schwarzschild black hole. In Schwarzschild coordinates, the metric takes the form (with $c=1$)
	\begin{equation}
		ds^2 = -\left( 1-\frac{r_h}{r}\right)d  t^2  + \left( 1-\frac{r_h}{r}\right)^{-1}d  r^2 + r^2d\Omega^2\,,
	\end{equation}
	where the Schwarzschild radius $r_h = 2GM$ is given in terms of Newton's constant $G$ and the mass of the black hole $M$. Using the conventions of~\cite{Bagchi:2021ban}, we now change coordinates according to $ r = {r_h} + \frac{1}{r_h}\epsilon\mathcal{r}^2$, where $\epsilon$ is a dimensionless parameter. This leads to the metric
	\begin{equation}
		\label{eq:schwarzschild-exp}
		\begin{split}
			ds^2 &= r_h^2d\Omega^2 + \epsilon\left[ -\frac{\mathcal{r}^2}{r_h^2}dt^2 + 4d\mathcal{r}^2 + 2\mathcal{r}^2d\Omega^2 \right] + \mathcal{O}(\epsilon^2)\,.
		\end{split}
	\end{equation}
	If we compare this with the Carroll expansion in~\eqref{eq:carroll-exp}, with $\epsilon$ playing the role of $c^2$, we may write
	\begin{equation}
		\begin{split}
			g_{\mu\nu} &= h_{\mu\nu} + \epsilon\eta_{AB}\tau_\mu^A\tau_\nu^B + \epsilon\Phi_{\mu\nu} + \mathcal{O}(\epsilon^2)\,,
		\end{split}
	\end{equation}
	where
	\begin{equation}
		\begin{split}
			h_{\mu\nu}dx^\mu dx^\nu &= r_h^2d\Omega^2\,,~~ \Phi_{\mu\nu} dx^\mu dx^\nu = 2\mathcal{r}^2d\Omega^2\\
			\eta_{AB}\tau_\mu^A\tau_\nu^B dx^\mu dx^\nu &= -\frac{\mathcal{r}^2}{r_h^2}dt^2 + 4d\mathcal{r}^2 \,.
		\end{split}
	\end{equation}
	The longitudinal space is $(1+1)$-dimensional Rindler space and the transverse space is a sphere of radius $r_h$. 
	
	\medskip
	
	\noindent \textbf{Near-horizon Polyakov string.}
	We will now use our expansion above to study closed bosonic string theory in the near-horizon region of a 4d Schwarzschild black hole. The Polyakov action is given by 
	\begin{equation}
		\label{eq:polyakov-lagrangian}
		S_{\text{P}} = - \frac{T}{2}\int d^2\sigma \sqrt{-\gamma}\gamma^{\alpha\beta} \D_\alpha X^\mu \D_\beta X^\nu g_{\mu\nu}(X)\, .
	\end{equation}
	Formally, the string Carroll expansion in terms of $\epsilon$ is equivalent to the approach presented in~\cite{Hartong:2021ekg,Hartong:2022dsx}, though we expand in powers of $\epsilon\sim c^2$ rather than $1/c^2$; in other words, it is an ultralocal, rather than a nonrelativistic, expansion. The expansions of the embedding scalars and the worldsheet metric are
	\begin{equation*}
		X^\mu = x^\mu + \epsilon y^\mu +\mathcal{O}(\epsilon^2)\,,\,  \gamma_{\alpha\beta} = \gamma_{(0)\alpha\beta} + \epsilon \gamma_{(2)\alpha\beta} +\mathcal{O}(\epsilon^2)\,,
	\end{equation*}
	where $\gamma_{(0)\alpha\beta}$ is Lorentzian.
	Hence, the expansion of the pullback of the target space metric is
	\begin{equation*}
		g_{\alpha\beta}(X) = \D_\alpha X^\mu \D_\beta X^\nu g_{\mu\nu}(X) = h_{\alpha\beta}(x) + \epsilon \hat\Phi_{\alpha\beta}(x,y) + \mathcal{O}(\epsilon^2)\,,
	\end{equation*}
	where
	\begin{equation}
		\begin{split}
			\hat\Phi_{\alpha\beta}(x,y) &= \tau_{\alpha\beta}(x) + \Phi_{\alpha\beta}(x) \\
			&\hspace{-1cm}+  2h_{\mu\nu}(x)\D_{(\alpha} x^\mu \D_{\beta)}y^\nu+ \D_\alpha x^\mu \D_\beta x^\nu y^\rho \D_\rho h_{\mu\nu}(x)\,. 
		\end{split}
	\end{equation}
	Here $\tau_{\alpha\beta}$ and $\Phi_{\alpha\beta}$ are the pullbacks of $\tau_{\mu\nu}$ and $\Phi_{\mu\nu}$ using the LO embedding scalars $x^\mu$.
	The Polyakov Lagrangian~\eqref{eq:polyakov-lagrangian} consequently acquires the $\epsilon$-expansion
	\begin{equation}
		\mathcal{L}_{\text{P}} = \mathcal{L}_{\text{P-LO}} + \epsilon\mathcal{L}_{\text{P-NLO}} +\mathcal{O}(\epsilon^2)\,.
	\end{equation}
	The LO Polyakov action is given by~\footnote{There are usually two different ways of taking Carroll limits of theories: the electric and magnetic limits, see, e.g.,~\cite{henneaux:2021yzg,deBoer:2021jej}. The magnetic Carroll \textit{particle} is characterised by having vanishing energy to leading order~\cite{deBoer:2021jej}. The string Carroll analogue of this is the vanishing of the Noether charges of longitudinal translations to leading order. Since the LO Polyakov action~\eqref{eq:LO-Polyakov} only probes the transverse space, these LO longitudinal Noether charges precisely vanish, implying that we're dealing with the magnetic Carroll string. We will introduce notions of electric and magnetic strings in upcoming work~\cite{UpcomingCarrollString}.}
	\begin{equation}
		\label{eq:LO-Polyakov}
		\mathcal{L}_{\text{P-LO}} = - \frac{T}{2}\sqrt{-{\gamma_{(0)}}}\gamma^{\alpha\beta}_{(0)} h_{\alpha\beta}(x)\,.
	\end{equation}
	The equation of motion for the embedding fields $x^\mu$ is
	\begin{equation}
		\label{eq:LO-x-eom}
		\begin{split}
			&\frac{1}{\sqrt{-{\gamma_{(0)}}}}\D_\alpha\left( \sqrt{-{\gamma_{(0)}}} \gamma_{(0)}^{\alpha\beta}\D_\beta x^\mu h_{\mu\nu}(x) \right)\\
			&\qquad= \frac{1}{2}\gamma^{\alpha\beta}_{(0)}\D_\alpha x^\rho \D_\beta x^\mu \D_\nu h_{\rho\mu}(x)\,,
		\end{split}
	\end{equation}
	while the LO Virasoro constraints, which are obtained by varying $\gamma_{(0)\alpha\beta}$, take the form
	\begin{equation}
		\label{eq:LO-Virasoro}
		T^{(0)}_{\alpha\beta} = h_{\alpha\beta} - \frac{1}{2}\gamma^{\gamma\delta}_{(0)}h_{\gamma\delta}\gamma_{(0)\alpha\beta} = 0\,.
	\end{equation}
	The next-to leading (NLO) Lagrangian is
	{\small\begin{equation}
			\label{eq:P-NLO-lagrangian}
			\begin{split}
				\mathcal{L}_{\text{P-NLO}} &= - \frac{T}{2} \sqrt{-{\gamma_{(0)}}} \left( \gamma_{(0)}^{\alpha\beta}\hat\Phi_{\alpha\beta}(x,y) - \frac{1}{2}G_{(0)}^{\alpha\beta\gamma\delta} h_{\alpha\beta}(x)\gamma_{(2)\gamma\delta} \right)\,,
			\end{split}
	\end{equation}}
	where we introduced the Wheeler--DeWitt metric
	\begin{equation}
		G_{(0)}^{\alpha\beta\gamma\delta} = \gamma_{(0)}^{\alpha\gamma}\gamma_{(0)}^{\delta\beta} + \gamma_{(0)}^{\alpha\delta}\gamma_{(0)}^{\gamma\beta}- \gamma_{(0)}^{\alpha\beta}\gamma_{(0)}^{\gamma\delta}\,.
	\end{equation}
	Varying $\gamma_{(2)\alpha\beta}$ reproduces the LO Virasoro constraints~\eqref{eq:LO-Virasoro}, while $\gamma_{(0)\alpha\beta}$ now imposes the NLO Virasoro constraints
	\begin{equation*}
		T^{(2)}_{\alpha\beta} = \hat\Phi_{\alpha\beta}(x,y) - \frac{1}{2}\gamma^{\gamma\delta}_{(0)}\hat\Phi_{\gamma\delta}(x,y)\gamma_{(0)\alpha\beta} + \text{terms w/ $\gamma_{(2)}$} = 0\,.
	\end{equation*}
	As we will see below, we may gauge fix $\gamma_{(2)\alpha\beta} = 0$, so we refrain from writing out those terms explicitly. The equation of motion for $y^\mu$ imposes the LO equation of motion for $x^\mu$~\eqref{eq:LO-x-eom}, while the equation for $x^\mu$ gives an equation for $y^\mu$.
	
	\medskip
	
	\noindent \textbf{Worldsheet symmetries.}
	We now investigate the worldsheet symmetries of the action we have constructed for our near-horizon string. It is instructive to contrast our construction with the null string~\cite{Isberg:1993av,Bagchi:2015nca}, which is forced to move on a null hypersurface in Lorentzian spacetime, so that the worldsheet symmetries become conformal Carrollian~\cite{Duval:2014uva,Duval:2014lpa} instead of remaining two copies of the Virasoro algebra. It was shown earlier in~\cite{Bagchi:2020ats, Bagchi:2021ban} that when a relativistic string moves in a near-horizon Rindler geometry, a Rindler structure is induced on the worldsheet, and conformal Carrollian structures appear when the string hits the horizon and becomes null. Since the expansion that we developed above does not lead to a null string, we expect the Virasoro structure to remain unaffected. We confirm this below. 
	
	The worldsheet metric $\gamma_{\alpha\beta}$ transforms infinitesimally under two-dimensional diffeomorphisms generated by $\xi^\alpha$ and local Weyl rescalings $\omega$ as
	\begin{equation}
		\delta\gamma_{\alpha\beta} = \pounds_\xi \gamma_{\alpha\beta} + 2 \omega \gamma_{\alpha\beta}\,.
	\end{equation}
	Like the worldsheet metric itself, these gauge parameters are expanded according to
	\begin{equation*}
		\xi^\alpha = \xi^\alpha_{(0)} + \epsilon \xi_{(2)}^\alpha + \mathcal{O}(\epsilon^2)\,,\qquad \omega = \omega_{(0)} + \epsilon \omega_{(2)} + \mathcal{O}(\epsilon^2)\,.
	\end{equation*}
	This means that $\gamma_{(0)\alpha\beta}$ and $\gamma_{(2)\alpha\beta}$ transform as 
	\begin{equation}
		\begin{split}
			\delta \gamma_{(0)\alpha\beta} &= \pounds_{\xi_{(0)}}\gamma_{(0)\alpha\beta} + 2\omega_{(0)}\gamma_{(0)\alpha\beta} \,,\\   
			\delta \gamma_{(2)\alpha\beta} &= \pounds_{\xi_{(0)}}\gamma_{(2)\alpha\beta} + \pounds_{\xi_{(2)}}\gamma_{(0)\alpha\beta}\\
			&\quad + 2\omega_{(2)}\gamma_{(0)\alpha\beta} + 2\omega_{(0)}\gamma_{(2)\alpha\beta} \,.
		\end{split}
	\end{equation}
	The relativistic embedding scalars $X^\mu$ transform as $\delta X^\mu = \xi^\alpha \D_\alpha X^\mu$ under $\xi^\alpha$, implying that $x^\mu$ and $y^\mu$ transform according to
	\begin{equation}
		\label{eq:action-of-diffeos}
		\delta x^\mu = \xi^\alpha_{(0)}\D_\alpha x^\mu\,,\qquad \delta y^\mu = \xi^\alpha_{(2)}\D_\alpha x^\mu + \xi^\alpha_{(0)}\D_\alpha y^\mu\,.
	\end{equation}
	We can use the three parameters $\xi_{(0)}^\alpha$ and $\omega_{(0)}$ to (locally) go to flat or conformal gauge
	\begin{equation}
		\label{eq:gauge-fixed-LO-WS-metric}
		\gamma_{(0)\alpha\beta} = \eta_{\alpha\beta}\,.
	\end{equation}
	This leaves us with residual diffeomorphism invariance consisting of those diffeomorphisms that can be undone by a Weyl transformation, i.e., $\pounds_{\xi_{(0)}}\eta_{\alpha\beta} + 2\omega_{(0)}\eta_{\alpha\beta} = 0$. We change to lightcone coordinates $\sigma^\pm = \sigma^0 \pm \sigma^1$, which leads to the following nonzero components of the gauge-fixed worldsheet metric~\eqref{eq:gauge-fixed-LO-WS-metric} and its inverse
	\begin{equation*}
		\eta_{-+} = \eta_{+-}=-\frac{1}{2}\, , \eta^{-+} = \eta^{+-} = -2\,,
		\text{with} \, \sqrt{-{\gamma_{(0)}}} = \frac{1}{2}\,.
	\end{equation*}
	The derivatives are $\D_\pm = \frac{1}{2} (\D_0 \pm \D_1)$. In lightcone coordinates, the residual diffeomorphisms become
	\begin{equation}
		\xi_{(0)} = \xi^-_{(0)}(\sigma^-)\D_- + \xi^+_{(0)}(\sigma^+) \D_+\,,
	\end{equation}
	where, since we are dealing with closed strings, $\xi^\pm_{(0)}(\sigma^\pm)$ are \textit{periodic} functions of their arguments. At NLO, we can choose $\xi_{(2)}$ such that $\gamma_{(2)\alpha\beta} = 0$ which leaves the \textit{same} residual gauge transformations that we had at LO, i.e., $\xi_{(2)} = \xi^-_{(2)}(\sigma^-)\D_- + \xi^+_{(2)}(\sigma^+) \D_+$. In other words, the algebra generated by residual gauge transformations will always be two copies of the Witt algebra. This is in keeping with the idea that since the background has two longitudinal directions (the 2d Rindler space with an $\epsilon$ in front of it, cf.~Eq.~\eqref{eq:schwarzschild-exp}), the string will orient itself along these directions so as to keep its relativistic worldsheet structure intact. 
	
	\medskip

	\noindent \textbf{The string freezes on the $S^2$.}
	Now we investigate the equations of motion and its leading order solutions. On a Schwarzschild background, the gauge-fixed LO Lagrangian becomes 
	\begin{equation}
		\begin{split}
			\mathcal{L}_{\text{P-LO}} &= -\frac{T r_h^2}{4}\eta^{\alpha\beta}( \D_\alpha x^\theta\D_\beta x^\theta + \D_\alpha x^\phi \D_\beta x^\phi \sin^2(x^\theta) )\\
			&= r_h^2 T ( \D_- x^\theta \D_+ x^\theta + \sin^2(x^\theta) \D_- x^\phi \D_+ x^\phi )  \,.
		\end{split}
	\end{equation}
	The nontrivial Virasoro constraints are 
	\begin{equation}
		\begin{split}
			0 &= T^{(0)}_{--} = (\D_- x^\theta)^2 + (\sin(x^\theta)  \D_- x^\phi )^2\,,\\
			0 &= T^{(0)}_{++} = ( \D_+ x^\theta)^2 + (\sin(x^\theta) \D_+ x^\phi )^2\,.
		\end{split}
	\end{equation}
	These are sums of squares, and thus, since the embedding scalars are real-valued, they are independent of $\sigma^\pm$; in other words
	\begin{equation}
		\label{eq:freezing-on-the-S2}
		x^\theta = x_0^\theta\,,\qquad x^\phi = x_0^\phi\,,
	\end{equation}
	where $x_0^\theta,x^\phi_0$ are constants. The equations of motion~\eqref{eq:LO-x-eom} for $x^\theta$ and $x^\phi$ take the form
	\begin{equation}
		\begin{split}
			\eta^{\alpha\beta} \D_\alpha \D_\beta x^\theta &= \cos(x^\theta)\sin(x^\theta)\eta^{\alpha\beta}\D_\alpha x^\phi \D_\beta x^\phi\,,\\
			0 &= \eta^{\alpha\beta}\D_\alpha\D_\beta(\sin^2(x^\theta) x^\phi)\,,
		\end{split}
	\end{equation}
	and are trivially solved by~\eqref{eq:freezing-on-the-S2}. We may understand the triviality of the solution as a consequence  of the fact that the equations of motion identify the worldsheet geometry with the LO two-dimensional target geometry. Since the former is Lorentzian and the latter Euclidean, they cannot be identified and so the solution must be trivial (cf.~Eq.~\eqref{eq:LO-Virasoro}).

	\medskip
	
	Let us pause to ponder the consequences of this. Our analysis has revealed that the string {\emph{freezes}} on the $S^2$, i.e., to leading order, the string is just a point on the sphere in the target spacetime. This could mean that the string is freely falling into the black hole, and, as viewed by a stationary observer at infinity, contracts to a point (see Fig.~\ref{fig:NH-Carroll-string}). There is another tantalising scenario where the string orients itself radially and hence a closed string could fold on itself. We discuss this briefly in the conclusions. 
	
	\noindent\textbf{NLO theory: the Rindler string.} As explained in the previous section, the string freezes on the sphere to leading order. We now investigate the properties of the NLO theory, where the longitudinal Rindler space will play an important role. The NLO Lagrangian in the near-horizon region of the Schwarzschild black hole is 
	\begin{align}
		\mathcal{L}_{\text{P-NLO}} 
		&=T \bigg[ - \frac{1}{r_h^2}(x^{\mathcal{r}})^2\D_-x^t  \D_+ x^t + 4 \D_- x^{\mathcal{r}} \D_+ x^{\mathcal{r}}\nn\\\nn
		&\quad+ 2(x^{\mathcal{r}})^2 (\D_- x^\theta \D_+ x^\theta + \sin^2(x^\theta) \D_- x^\phi \D_+ x^\phi ) \bigg]\\
		&\quad+ y^\mu \frac{\delta \mathcal{L}_{\text{P-LO}}}{\delta x^\mu}\,.
	\end{align}
	After imposing the LO equations of motion and the LO Virasoro constraints, which lead to~\eqref{eq:freezing-on-the-S2}, the equations for $x^t$ and $x^{\mathcal{r}}$ become
	\begin{equation}\label{eq:NLO-eqns-for-x}
		\begin{split}
			0 &= x^{\mathcal{r}}( \D_- x^{\mathcal{r}}\D_+ x^t + \D_+ x^{\mathcal{r}}\D_- x^t ) + (x^{\mathcal{r}})^2 \D_-\D_+ x^t\,,\\
			0 &= \frac{1}{r_h^2}x^{\mathcal{r}}\D_- x^t \D_+ x^t + 4\D_-\D_+ x^{\mathcal{r}}\,.
		\end{split}
	\end{equation}
	The NLO Virasoro constraints are $T^{(2)}_{--} = T^{(2)}_{++} = 0$, which, after using~\eqref{eq:freezing-on-the-S2}, read
	\begin{equation}
		\label{eq:NLO-constraints}
		\begin{split}
			0 &= \tau_{--} = - \frac{{x^\mathcal{r}}^2}{r_h^2}(\D_- x^t)^2 + 4(\D_- x^{\mathcal{r}})^2\,,\\
			0 &= \tau_{++} = - \frac{{x^\mathcal{r}}^2}{r_h^2}( \D_+ x^t)^2 + 4(\D_+ x^{\mathcal{r}})^2\,.
		\end{split}
	\end{equation}
	In addition, there are the equations of motion of $x^\theta$ and $x^\phi$ which after imposing~\eqref{eq:freezing-on-the-S2} read
	\begin{equation}\label{eq:NLO-eqns-for-y}
		\D_- \D_+ y^\theta = \D_-\D_+ y^\phi = 0\,,
	\end{equation}
	which are standard 2d wave equations. 
	
	Again, we pause to consider the consequences of our findings. Once the string freezes on the sphere, the 2d Rindler space which was subleading in the string Carroll expansion of the metric~\eqref{eq:schwarzschild-exp}, becomes important and  the equations in \eqref{eq:NLO-eqns-for-x} provide us with information about how the string sees this 2d spacetime. We will have a more intuitive way of understanding this below. Although the string freezes on the $S^2$ to leading order, there are NLO fluctuations on the sphere which must be taken into account at the same order as the Rindler equations. This leads to  
	\eqref{eq:NLO-eqns-for-y}, which are standard wave equations and tell us that there are standard string excitation modes arising from the subleading fluctuations on the sphere. The spherical coordinates thus play an important role in retaining string-like features. 
	
	\medskip
	
	\noindent \textbf{String as a massless Rindler particle.}
	We will now present a solution to the equations of motion of the action we described above that will make the physical picture of strings probing black hole near-horizon geometries very interesting.  
	A solution to the NLO equations of motion \eqref{eq:NLO-eqns-for-x}, \eqref{eq:NLO-eqns-for-y} that satisfies the NLO Virasoro constraints \eqref{eq:NLO-constraints} is
	\begin{align}
		x^t &= x^t_0 \pm  {r_h}\log(2\sigma^0), \quad x^{\mathcal{r}} = r_h\sqrt{2\sigma^0}, \\
		y^{\theta} &= y^{\theta}_+(\sigma^+) + y^{\theta}_-(\sigma^-), \quad y^{\phi} = y^{\phi}_+(\sigma^+) + y^{\phi}_-(\sigma^-)\,.\nn
	\end{align}
	The solutions for the subleading embedding fields on the $S^2$ admit standard mode expansions, while the worldline on the longitudinal 2d Rindler space can be written as 
	\begin{equation}
		\label{eq:rindler-geodesic}
		x^t = x_0^t \pm {2r_h}\log(x^{\mathcal{r}}/r_h)\,.
	\end{equation}
	The above is nothing but a (infalling/outgoing) null geodesic of 2d Rindler space. To obtain these solutions, note that a null geodesic in Rindler space must satisfy the relation $\tau_{\mu\nu} \dot x^\mu \dot x^\nu = 0$, where the dot denotes a derivative with respect to an appropriately chosen affine parameter. Solving this leads to~\eqref{eq:rindler-geodesic}, which also satisfies the geodesic equation on Rindler space.
	
	Hence we find that a string which freely falls into the 4d Schwarzschild black hole at LO (in terms of the embedding fields), when viewed by a stationary asymptotic observer, shrinks to a point as it enters the near-horizon region. This point particle then moves along a null geodesic of the 2d Rindler space. At NLO there are  wave-like excitations for the subleading sphere coordinates $y^{\theta}$, $y^{\phi}$ giving the string its stringy nature. We expect this feature to be present for any generic non-extremal black hole with spherical topology in arbitrary dimensions.

	\medskip

	\medskip
	
	\noindent \textbf{Concluding remarks.} In this paper, we have taken initial steps towards understanding the fate of strings approaching a black hole. In the near-horizon geometry of a 4d Schwarzschild black hole, the string encounters a 2d Rindler spacetime together with a parametrically large sphere. This geometry takes the form of a string Carroll expansion, where the 2d Rindler space forms a Lorentzian submanifold. We expanded the closed bosonic string action in terms of this novel geometric structure and uncovered a theory that determines the classical dynamics of the string in the near-horizon region. We found that the Virasoro symmetries are still intact as residual symmetries of the worldsheet. Interestingly, the solutions of the leading order equations of motion show that the string freezes on the leading order sphere, which means that the string shrinks to a point in this limit.  
	
	There is another intriguing possibility: the string may orient itself radially without any transverse extent while entering the sphere so as to respect the leading order equations of motion while retaining its dimensions. In this case, if we start out with closed strings, the process seems to indicate that the string would fold onto itself as it goes into the near-horizon region so that it does not have any angular extent to leading order. This is reminiscent of strings that fold when their tension goes to zero \cite{Bagchi:2021ban}, and also of earlier literature where folded strings appear near black holes \cite{Bars:1994sv}. We will return to this very interesting story in future work.

	As the string freezes on the $S^2$, the subleading 2d Rindler becomes fundamentally important to the near-horizon string. We find that the equations of motion constrain the string to move along null geodesics of this 2d subspace. Although our analysis explicitly considered the dynamics of strings in the near-horizon geometry of four-dimensional Schwarzschild black holes, our analysis is expected to generalise in a straightforward way to strings probing any non-extremal blackhole and the features we have discussed would hold in general.  
	
	There are many questions and directions for further investigation, the foremost amongst which is the question of what happens when we quantise this near-horizon string. Does such a quantum theory describe a universal sector of the quantum superstring defined on a weakly curved part (so as to keep the $\alpha'$ corrections under control) of a 10-dimensional background that contains a 4-dimensional Schwarzschild geometry. In this work we focused on stationary observers at infinity. An important follow-up question is what would happen if we use coordinates that are regular near the horizon (like infalling Eddington--Finkelstein coordinates). What geometry does the string perceive as it closes in on the horizon? Since the latter is a null hypersurface it would be conceivable that this is again some string Carroll geometry.

	The detailed phase space analysis is work to appear soon~\cite{UpcomingCarrollString}. There we will discuss two Carroll limits of the relativistic string and show that the string considered in this letter is the magnetic Carroll string. It would be interesting to understand if the electric Carroll string is important for strings near black holes as well.  
	
	\medskip
	
	\noindent \textbf{Acknowledgements.} We are grateful to Daniel Grumiller, Niels Obers, and Joan Sim\'on for useful discussions. AB is partially supported by a Swarnajayanti Fellowship from the Science and Engineering Research Board
	(SERB) under grant SB/SJF/2019-20/08 and also by SERB grant CRG/2020/002035. ArB is supported in part by an OPERA grant and a seed grant NFSG/PIL/2023/P3816 from BITS-Pilani. JH was supported by the Royal Society University Research Fellowship
	Renewal “Non-Lorentzian String Theory” (grant number URF\textbackslash
	R\textbackslash 221038). AB and JH were also supported by the Royal Society International Exchange Grant `Carrollian Symmetry and String Theory' (IES\textbackslash R3\textbackslash 213165). The work of EH is supported by Villum Foundation Experiment project 00050317, ``Exploring the wonderland of Carrollian physics''. KK was partially supported by the SERB grant SB/SJF/2019-20/08. KK and MM thank the hospitality of the University of Edinburgh while this work was in progress, and this visit was supported by the Royal Society International Exchange grant (IES\textbackslash R3\textbackslash 213165).

	\providecommand{\href}[2]{#2}\begingroup\raggedright\endgroup

\end{document}